\documentclass[english,aip,pop,reprint]{revtex4-2}
\usepackage{ae,aecompl}
\usepackage[T1]{fontenc}
\usepackage[latin9]{inputenc}
\usepackage{color}
\usepackage{amsmath}
\usepackage{amssymb}
\usepackage{graphicx}

\makeatletter

\newcommand{\lyxdot}{.}

\makeatother

\usepackage{babel}
\begin{document}
\title{Three-dimensional plasmoid-mediated reconnection and turbulence in Hall magnetohydrodynamics}
\author{Yi-Min Huang}
\email{yiminh@princeton.edu}

\affiliation{Department of Astrophysical Sciences, Princeton University, Princeton, New Jersey 08540, USA}
\author{Amitava Bhattacharjee}
\affiliation{Department of Astrophysical Sciences, Princeton University, Princeton, New Jersey 08540, USA}
\begin{abstract}

Plasmoid instability accelerates reconnection in collisional plasmas by transforming a laminar reconnection layer into numerous plasmoids connected by secondary current sheets in two dimensions (2D) and by fostering self-generated turbulent reconnection in three dimensions (3D). In large-scale astrophysical and space systems, plasmoid instability likely initiates in the collisional regime but may transition into the collisionless regime as the fragmentation of the current sheet progresses toward kinetic scales. Hall MHD models are widely regarded as a simplified yet effective representation of the transition from collisional to collisionless reconnection. However, plasmoid instability in 2D Hall MHD simulations often leads to a single-X-line reconnection configuration, which significantly differs from fully kinetic particle-in-cell simulation results. This study shows that single-X-line reconnection is less likely to occur in 3D compared to 2D. Moreover, depending on the Lundquist number and the ratio between the system size and the kinetic scale, Hall MHD can also realize 3D self-generated turbulent reconnection. We analyze the features of the self-generated turbulent state, including the energy power spectra and the scale dependence of turbulent eddy anisotropy.

\end{abstract}
\maketitle

\section{Introduction}

Magnetic reconnection is a fundamental process that alters the connectivity
of magnetic field lines, releasing stored magnetic energy. The magnetic energy released by reconnection is transformed into kinetic, thermal,
and non-thermal energy of plasma in explosive events in nature and
laboratories, such as geomagnetic substorms, solar flares, coronal
mass ejections (CMEs), gamma-ray bursts, and sawtooth crashes in fusion
plasmas.\citep{Biskamp2000,PriestF2000,ZweibelY2009,YamadaKJ2010,ZweibelY2016,PontinP2022,JiDJLSY2022} 

Magnetic reconnection occurs at current sheets. In recent years,
there has been growing evidence that large-scale reconnection current
sheets will likely become fragmented due to the plasmoid instability.\citep{LoureiroSC2007,BhattacharjeeHYR2009}
The plasmoid instability has been found in a wide range of plasma
models, including resistive magnetohydrodynamics (MHD), \citep{Biskamp1986,ShibataT2001,LoureiroSC2007,Lapenta2008,BhattacharjeeHYR2009,CassakSD2009,HuangB2010,BartaBKS2011,HuangBS2011,ShenLM2011,LoureiroSSU2012,NiZHLM2012,HuangB2012,HuangB2013,Takamoto2013,WyperP2014}
Hall MHD,\citep{ShepherdC2010,HuangBS2011,BaalrudBHG2011} and fully
kinetic particle-in-cell (PIC) simulations. \citep{DaughtonSK2006,DaughtonRAKYB2009,DaughtonRKYABB2011,FermoDS2012,DaughtonNKRL2014,StanierDLLB2019} 

High-resolution two-dimensional (2D) resistive MHD simulations indicate
that the plasmoid instability leads to a fast reconnection rate that
is nearly independent of the resistivity.\citep{BhattacharjeeHYR2009,HuangB2010,LoureiroSSU2012}
In three-dimensional (3D) reconnection with a guide field, in addition
to modes that are symmetric along the out-of-plane direction,
oblique tearing modes can become unstable.\citep{BaalrudBH2012} Interaction
of oblique tearing modes leads to self-generated turbulent reconnection
without the need for external forcing.\citep{DaughtonRKYABB2011,HuangB2016,YangLGLLHZF2020,BegRH2022} 

Taking into account two-fluid and kinetic physics, the plasmoid instability can trigger even faster collisionless reconnection if the fragmented current sheets reach a kinetic scale $\delta_{i}$ that corresponds to the ion skin depth $d_{i}$ or the ion sound Larmor radius $\rho_{s}$,
the latter being for cases with a strong guide field.\citep{DaughtonRAKYB2009,ShepherdC2010,HuangBS2011}
This effect leads to the consideration of reconnection ``phase diagram''
in the literature.\citep{JiD2011,HuangBS2011,HuangB2013,CassakD2013}
The phase diagram organizes various types of reconnection in the parameter
space of two dimensionless parameters: the Lundquist number $S\equiv LV_{A}/\eta$
and the system size to the kinetic scale ratio $\Lambda\equiv L/\delta_{i}$
. Here, $L$ is a characteristic length scale of the reconnection
layer along the outflow direction, $V_{A}$ is the Alfv\'en speed of the reconnecting magnetic
field component, and $\eta$ is the magnetic diffusivity. Within resistive
MHD, the reconnection current sheet becomes unstable when the Lundquist
number $S$ exceeds a critical value $S_{c}\sim10^{4}$.\citep{BhattacharjeeHYR2009,HuangCB2019}
Once the current sheet becomes unstable, the fragmentation continues
until the secondary current sheets become marginally stable. This
condition yields the typical widths of the secondary current sheets
to scale as $\delta_{c}\sim LS_{c}^{1/2}/S$.\citep{HuangB2010} Comparing
the secondary current sheet width $\delta_c$ with the kinetic scale
$\delta_{i}$ yields a heuristic criterion $\Lambda<S/S_{c}^{1/2}$
for the plasmoid-mediated onset of collisionless reconnection.

Based on simple estimates, we can see that this onset criterion is
satisfied for numerous types of reconnection events in the solar atmosphere
and astrophysical systems.\citep{JiD2011} For example, consider post-CME
current sheets in the corona\citep{GuoBH2013,LinMSRRZWL2015} and
UV (ultraviolet) bursts in the transition region.\citep{InnesGHB2015,YoungTP2018,GuoDHPB2020}
Typical lengths of post-CME current sheets are on the order of $\sim10^{9}\text{m}$,
Lundquist numbers $S\sim10^{14}$, and $d_{i}\sim1\text{m}$; Hence,
the ratio $\Lambda=L/d_{i}\sim10^{9}$, which is smaller than $S/S_{c}^{1/2}\sim10^{12}$.
For UV bursts, the lengths are estimated to be on the order of $10^{5}\text{m}$,
the Lundquist numbers $S\sim10^{10}$, and $d_{i}\sim0.1\text{m}$.
The ratio $\Lambda\sim10^{6}$ is smaller than $S/S_{c}^{1/2}\sim10^{8}$.
Therefore, plasmoid-mediated onset of collisionless reconnection is
potentially important for both types of reconnection events.

Evidence of plasmoids in both types of reconnection events is abundant. In post-CME current sheets, plasmoid-like structures have been regularly observed
as moving blobs. \citep{LinCF2008,GuoBH2013,LinMSRRZWL2015} In UV
bursts, although plasmoids are not directly visible, their existence
may be inferred from the shapes of emission line profiles, which provide information on reconnection outflow through the Doppler shift. The emission line profiles of UV bursts typically have a triangle shape, which is consistent with a highly fluctuating reconnection outflow associated with plasmoids. In contrast, a single-X-line reconnection geometry will predict a double-peak line profile, which is inconsistent with observation.\citep{InnesGHB2015,GuoDHPB2020} 

To model the transition from collisional to collisionless
reconnection mediated by the plasmoid instability, particle-in-cell (PIC) simulation provides a first-principles, fully
kinetic description.\citep{DaughtonRAKYB2009,StanierDLLB2019} However, PIC
simulations, especially in 3D, are usually limited to relatively small
system sizes due to the computational cost. To simulate systems of larger
sizes, this study employs a Hall MHD model incorporating two-fluid effects through the Hall terms in the generalized Ohm's law. Hall MHD is often
regarded as a minimal model that captures important aspects of collisonless
or weakly collisional reconnection beyond resistive MHD.\citep{BirnDSRDHKMBOP2001}
 Previous 2D Hall MHD studies have demonstrated the onset
of Hall reconnection mediated by plasmoid instability.\citep{ShepherdC2010,HuangBS2011} 

From the perspective of plasmoid formation, 2D Hall MHD reconnection is somewhat anomalous: it tends to settle to a single-X-line geometry after expelling all the plasmoids from the reconnection layer. This feature makes it qualitatively different from fully kinetic PIC simulations, where new plasmoids are constantly generated.\citep{DaughtonRAKYB2009} The single-X-line geometry of Hall reconnection also appears to be inconsistent with observations of post-CME current sheets and UV bursts, which indicate that plasmoids continue to form throughout the entire period of the events. Although non-single-X-line 2D Hall reconnection has been demonstrated, it appears to require a fairly large system size ($L/d_{i} \gtrsim 5000$) and may only exist in a narrow parameter space of the phase diagram.\citep{HuangBS2011}

The observed persistence of plasmoid formation throughout solar reconnection events raises an important question regarding whether Hall MHD can adequately describe these phenomena.  However, it is important to note that the preference for single-X-line reconnection in Hall MHD may result from the assumed 2D symmetry. Will 3D systems allow the formation of complex 3D structures and significantly change the picture? Moreover, if 3D Hall MHD reconnection geometry deviates from a single X-line, can it support self-generated turbulent reconnection?

To address these questions, we perform 3D Hall MHD simulations of
reconnection and compare the results with the corresponding 2D simulations. We aim for large system sizes, up to $L/d_{i}=1000$. To mimic natural reconnection events, we start from an initial current sheet that is significantly thicker than $d_{i}$ and let it thin down self-consistently. After the plasmoid instability triggers the onset of Hall reconnection, we continue to follow the evolution until it reaches a saturated phase.

This paper is organized as follows. Section \ref{sec:Model} describes the governing equations and the simulation setup in detail. In Section \ref{sec:Geometry}, we present simulation results of reconnection geometry and reconnection rate for various 2D and 3D runs. Section \ref{sec:Characteristics-of-the-Turbulence} looks more deeply into the characteristics of the fully developed reconnection state by examining the energy power spectra and two-point structure functions of kinetic and magnetic fluctuations.  Finally, we discuss the implications of our findings for large-scale astrophysical reconnection and conclude in Section \ref{sec:Conclusion}.

\section{Governing Equations and Model Setup \label{sec:Model}}

\begin{figure}
\begin{centering}
\includegraphics[clip,width=1\columnwidth]{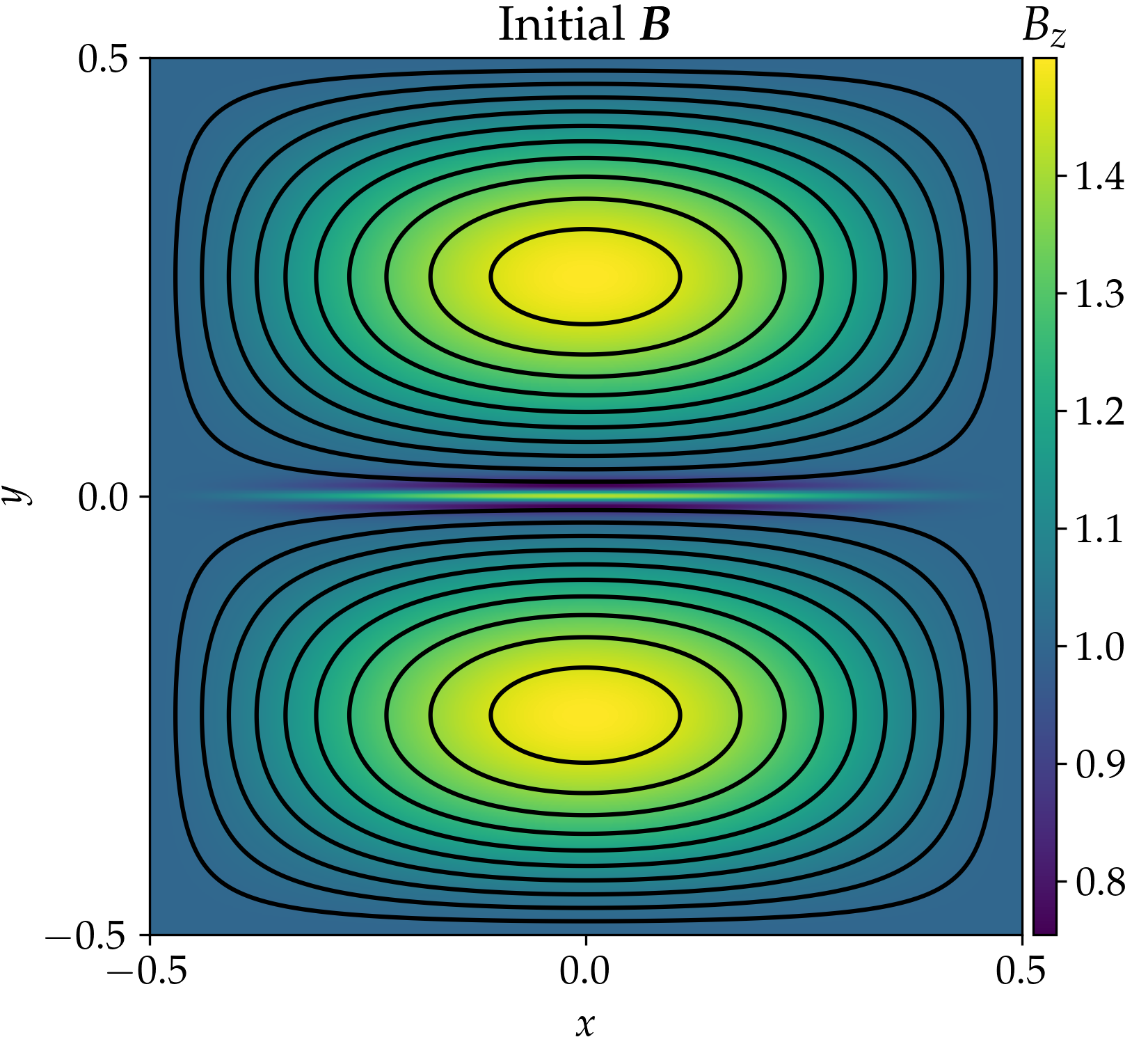}
\par\end{centering}
\caption{Initial magnetic field configuration. Black lines are stream lines
of the in-plane magnetic field component, and colormap shows the out-of-plane
component $B_{z}$.\label{fig:Initial-magnetic-field}}
\end{figure}

We employ the visco-resistive Hall magnetohydrodynamics model with
an adiabatic equation of state. In normalized units, the equations
are: 

\begin{equation}
\partial_{t}\rho+\nabla\cdot\left(\rho\boldsymbol{v}\right)=0,\label{eq:1}
\end{equation}
\begin{align}
\partial_{t}\left(\rho\boldsymbol{v}\right)+\nabla\cdot\left(\rho\boldsymbol{vv}\right)= & -\nabla\left(p+\frac{B^{2}}{2}\right)+\nabla\cdot\left(\boldsymbol{BB}\right)\nonumber \\
 & +\nabla\cdot\left(\rho\nu\frac{\nabla\boldsymbol{v}+\nabla\boldsymbol{v}^{T}}{2}\right),\label{eq:2}
\end{align}
\begin{equation}
\partial_{t}p+\nabla\cdot(p\boldsymbol{v})=-(\gamma-1)p\nabla\cdot\boldsymbol{v},\label{eq:3}
\end{equation}
\begin{equation}
\partial_{t}\boldsymbol{B}=-\nabla\times\boldsymbol{E},\label{eq:4}
\end{equation}
\begin{equation}
\boldsymbol{E=}-\boldsymbol{v}\times\boldsymbol{B}+d_{i}\frac{\boldsymbol{J}\times\boldsymbol{B}-\nabla p_{e}}{\rho}+\eta\boldsymbol{J}-\eta_{H}\nabla^{2}\boldsymbol{J}.\label{eq:Ohm}
\end{equation}
Here, standard notations are used: $\rho$
is the plasma density; $\boldsymbol{v}$ is the ion velocity; $p$
is the total plasma pressure; $p_{e}$ is the electron pressure; $\boldsymbol{B}$
is the magnetic field; $\boldsymbol{E}$ is the electric field; $\boldsymbol{J}=\nabla\times\boldsymbol{B}$
is the electric current density; $\eta$ is the resistivity; $\eta_{H}$
is the hyper-resistivity;\citep{HuangBF2013} $\nu$ is the viscosity;
and $d_{i}$ is the ion skin depth. For simplicity, we assume that the ions and electrons are locally in thermal equilibrium. Therefore,
the ion pressure $p_{i}$ and the electron pressure $p_{e}$ are equal,
and the total plasma pressure $p=p_{i}+p_{e}=2p_{e}$. Electron inertia
terms are neglected in the generalized Ohm's law, Eq. (\ref{eq:Ohm}).

The normalization of Eqs. (\ref{eq:1}) -- (\ref{eq:Ohm})
is based on constant reference values of the density $n_{0}$, and
the magnetic field $B_{0}$. Lengths are normalized to the system
size $L$, and time is normalized to the global Alfv\'en time $t_{A}=L/V_{A}$,
where $V_{A}=B_{0}/\sqrt{4\pi n_{0}m_{i}}$ and $m_{i}$ is the ion
mass. The normalization of physical variables is given by (normalized
variables $\to$ expressions in Gaussian units): $\rho\to\rho/n_{0}m_{i}$,
$\boldsymbol{B}\to\boldsymbol{B}/B_{0}$, $\boldsymbol{E}\to c\boldsymbol{E}/B_{0}V_{A}$,
$\boldsymbol{v}\to\boldsymbol{v}/V_{A}$, $p\to p/n_{0}m_{i}V_{A}^{2}$,
$\boldsymbol{J}\to\boldsymbol{J}/(B_{0}c/4\pi L)$, and $d_{i}\to d_{i}/L\equiv\sqrt{m_{i}c^{2}/4\pi n_{0}e^{2}}/L$. 

The simulation setup is similar to that employed in previous studies.\citep{HuangB2010,HuangBS2011,HuangB2012,HuangB2016,HuangCB2017}
In this setup, the coalescence of two magnetic flux tubes drives magnetic
reconnection. The simulation box is \textcolor{black}{a 3D cube in
the domain $(x,y,z)\in[-1/2,1/2]\times[-1/2,1/2]\times[-1/2,1/2]$}.
In normalized units, the initial magnetic field is
given by $\boldsymbol{B}=B_{z}\boldsymbol{\hat{z}}+\boldsymbol{\hat{z}}\times\nabla\psi$,
where $\psi=\tanh\left(y/h\right)\psi_{0}$ and $\psi_{0}=\cos\left(\pi x\right)\sin\left(2\pi y\right)/2\pi$.
Here, the parameter $h$ determines the initial thickness of the current
layer between the flux tubes. The out-of-plane magnetic field component
$B_{z}$ is non-uniform such that the initial configuration is approximately
force-balanced. Precisely, $B_{z}$ satisfies 
\begin{equation}
B_{z}^{2}=B_{z0}^{2}+5\pi^{2}\psi_{0}^{2}+\left(\partial_{y}\psi_{0}\right)^{2}-\left(\partial_{y}\psi\right)^{2},\label{eq:Bz}
\end{equation}
where the parameter $B_{z0}$ sets the guide field strength. In this
study, the guide field strength $B_{z0}$ is set to unity. The initial
current sheet thickness $h=0.01$. In the upstream region of the current
layer, the reconnecting component $B_{x}$ and the guide field $B_{z}$ are both approximately equal to unity. Figure \ref{fig:Initial-magnetic-field}
shows the initial magnetic field configuration.

The initial plasma density and pressure are both uniform, with
$\rho=1$ and $p_i=p_e=1$ in normalized units. The heat
capacity ratio $\gamma=5/3$ is assumed. Perfectly conducting and
free-slip boundary conditions are imposed along both $x$ and
$y$ directions. Specifically, we have $\boldsymbol{v}\cdot\hat{\boldsymbol{n}}=0$,
$\hat{\boldsymbol{n}}\cdot\nabla\left(\hat{\boldsymbol{n}}\times\boldsymbol{v}\right)=0$,
and $\boldsymbol{B}\cdot\hat{\boldsymbol{n}}=0$ on the boundaries.
Here, $\hat{\boldsymbol{n}}$ is the unit vector normal to the boundary.
The $z$ direction is assumed to be periodic. 

This model system is numerically solved with a massively
parallel HMHD code. The numerical algorithm \citep{GuzdarDMHL1993}
approximates spatial derivatives by finite differences with a five-point
stencil in each direction, augmented with a fourth-order numerical
dissipation for numerical stability. The time-stepping scheme can
be chosen from several options, including a second-order accurate trapezoidal
leapfrog method and various strong-stability-preserving Runge--Kutta
methods.\citep{GottliebST2001,SpiteriR2002} We employ the second-order
accurate trapezoidal leapfrog method in this study.

We perform two sets of simulations with the ion skin depth $d_{i}=0.002$
and $0.001$. The plasma resistivity and viscosity $\nu$ are both
set to a fixed value $5\times10^{-6}$. Using the box size as the
length scale $L$, the system size to the ion skin depth ratio $L/d_{i}$
is $500$ and $1000$, respectively. The Lundquist number $S=V_{A}L/\eta=2\times10^{5}$
and the magnetic Prandtl number $Pr_{m}\equiv\nu/\eta=1$. Because
Hall MHD tends to develop fine structures at small
scales, we add a small hyper-resistivity $\eta_{H}=10^{-13}$ to smooth
fluctuations at grid scales. The initial velocity is seeded with a random noise of amplitude $10^{-3}$ to trigger the plasmoid instability.
We perform 2D and 3D simulations with the same parameters and compare
the results. 

The 3D simulation mesh size is $N_{x}\times N_{y}\times N_{z}=2000\times1000\times2000$ ($2000\times1000$ for 2D),
where the grid sizes are uniform along both $x$ and $z$ directions,
and packed along the $y$ direction around the midplane to resolve
the reconnection layer better. The grid size along the $y$ direction
is $\Delta y=10^{-4}$ near the midplane ($y=0$). Moving away from
the midplane, the grid size gradually increases and reaches $\Delta y=0.005$ near the boundaries at $y=\pm1/2$.

\section{Reconnection Geometry and Reconnection Rate in Two-dimension and
Three-Dimension Simulations \label{sec:Geometry}}

\begin{figure}
\begin{centering}
\includegraphics[clip,width=1\columnwidth]{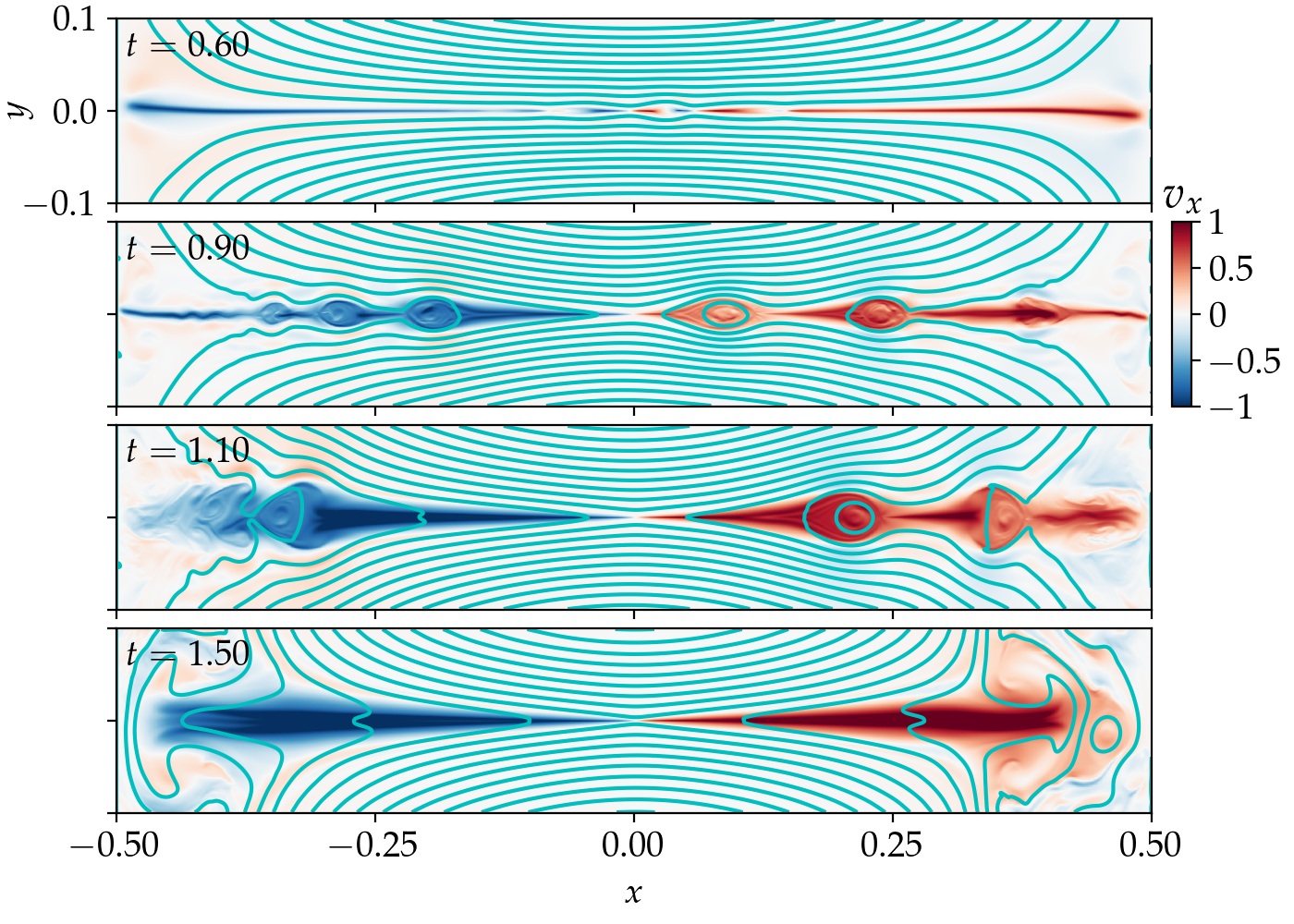}
\par\end{centering}
\caption{Time sequence for the 2D simulation with $d_{i}=0.002$ in the reconnection
layer. The colormap denotes the reconnection outflow velocity $v_{x}$
(multimedia view).\label{fig:2D_time_seq}}
\end{figure}

Figure \ref{fig:2D_time_seq} and the associate animation show the
time evolution of the 2D simulation with $d_{i}=0.002$, corresponding
to $L/d_{i}=500$. Here, the colormap denotes the reconnection outflow
velocity $v_{x}$. The initial thinning of the current sheet is accompanied by the formation of reconnection outflow jets. At around $t=0.6$, several plasmoids start to form in the reconnection layer. Subsequently, the onset of Hall reconnection expels all the plasmoids, and the reconnection
geometry settles to a single-X-line configuration at $t=1.5$. The
2D simulation with $d_{i}=0.001$ also exhibits qualitatively similar
behavior, as shown in the animation in the supplementary material.
These two runs demonstrate the strong tendency of 2D Hall MHD to form
a single-X-line reconnection geometry, similar to the results of
previous studies.\citep{ShepherdC2010,HuangBS2011} 

\begin{figure}
\includegraphics[width=1\columnwidth]{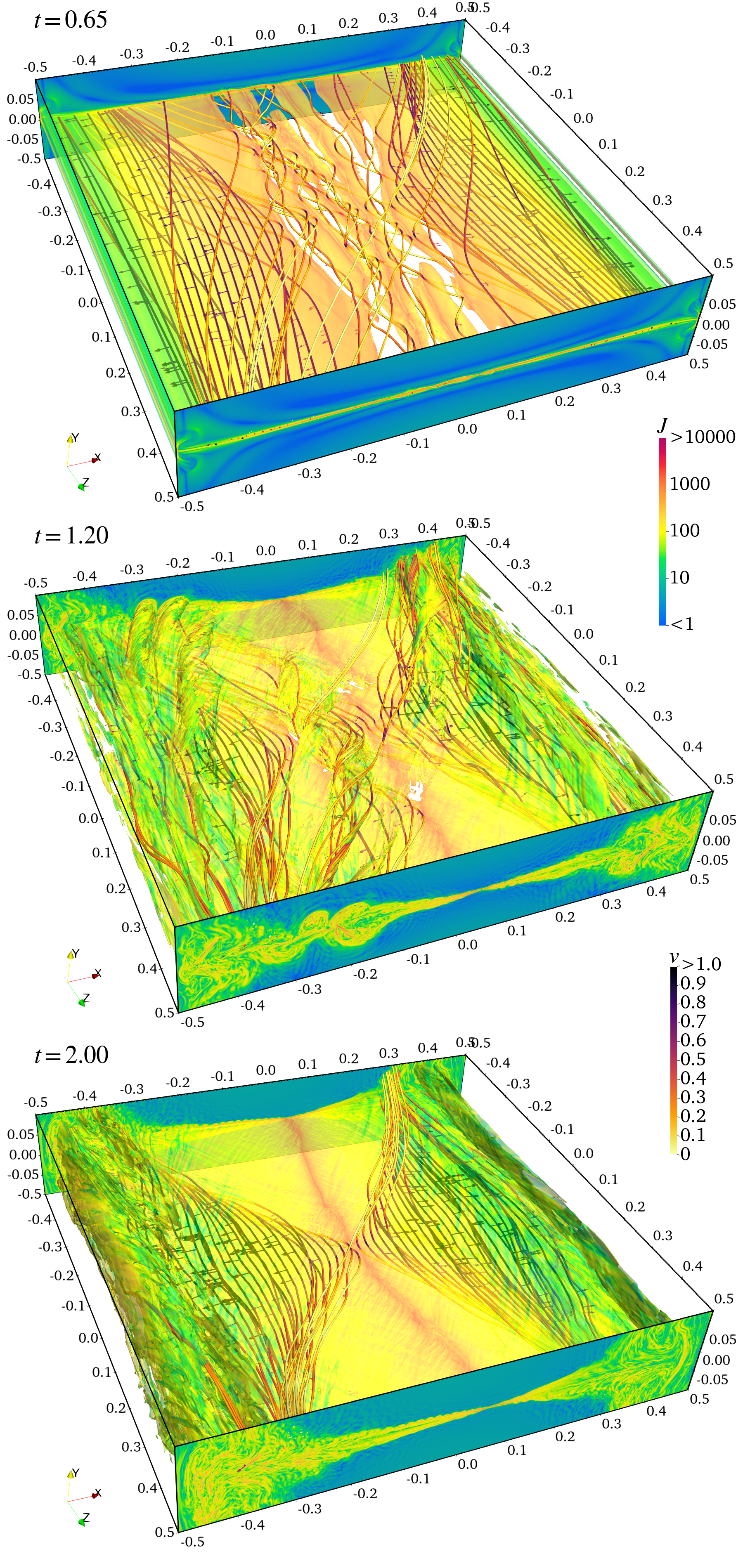}

\caption{Time sequence for the 3D simulation with $d_{i}=0.002$. Here, the
two end plates at $z=\pm0.5$ and the isosurfaces of the flow speed at $v=0.4$ are colored according to the magnitude of the current
density $J$ in the logarithmic scale. Samples of magnetic field lines
are colored according to the flow speed $v$. The arrows attached
to field lines denote the local velocity vector (multimedia view).
\label{fig:3D_time_seq_1} }
\end{figure}
\begin{figure}
\includegraphics[width=1\columnwidth]{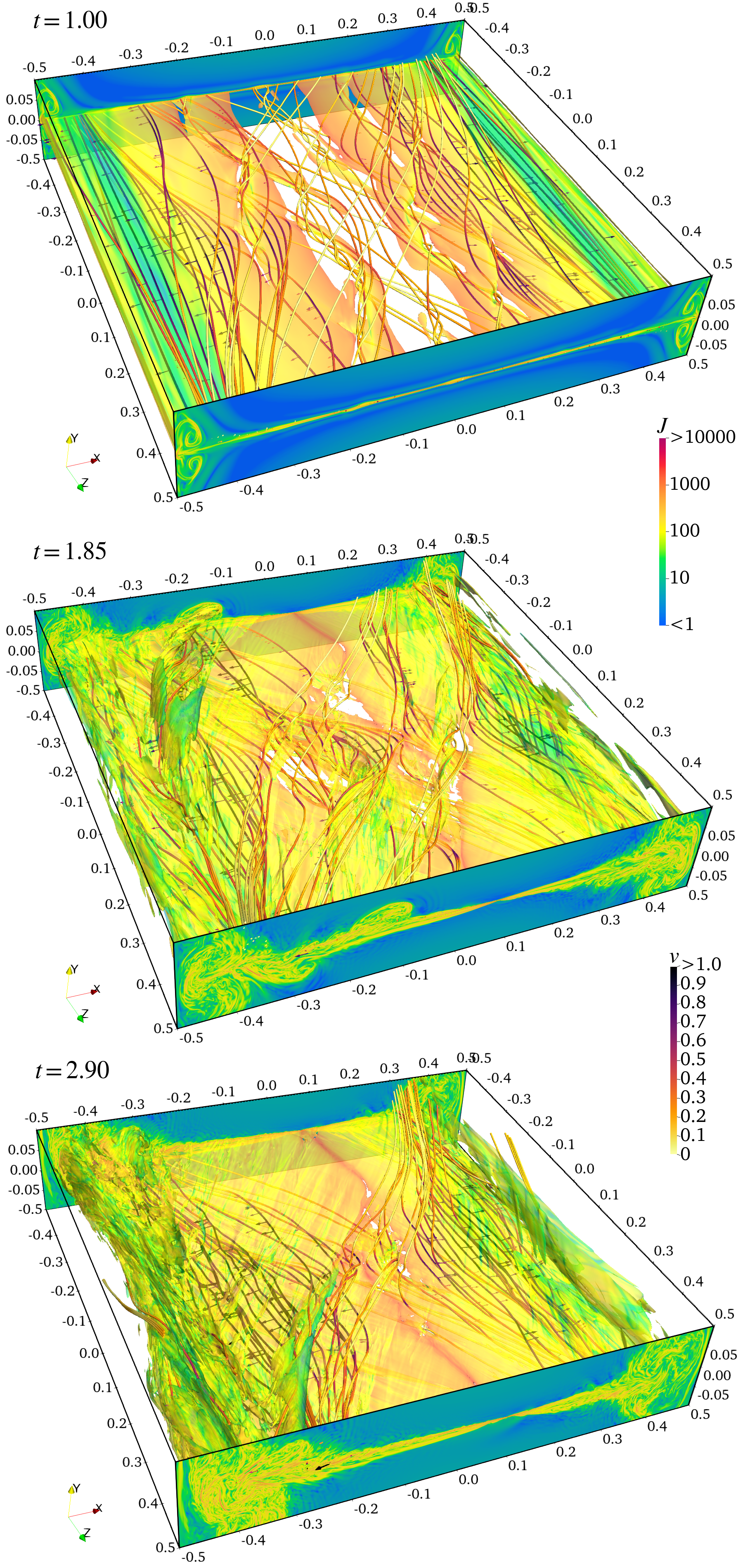}

\caption{Time sequence for the 3D simulation with $d_{i}=0.001$ (multimedia
view).\label{fig:3D_time_seq_2}}
\end{figure}

Three-dimensional (3D) simulations of the same parameters show different
behaviors. In the time evolution of the first case with $d_{i}=0.002$,
shown in Figure \ref{fig:3D_time_seq_1}, the thinning current sheet
first becomes unstable, forming flux ropes. The interaction between
flux ropes leads to complex dynamics and the formation of chaotic
field lines. However, all the flux ropes are eventually ejected and the magnetic field self-organizes into a nearly single-X-line
configuration. The final state is similar to its 2D counterpart, even
though a translational symmetry is not imposed. However, for the second
case with $d_{i}=0.001$ ($L/d_{i}=1000$), shown in Figure \ref{fig:3D_time_seq_2},
the magnetic configuration does not settle to a single X-line. Instead,
it appears to develop a self-generated turbulent state similar to
previous resistive MHD and PIC simulations. 

To quantify the 3D reconnection rate, we first average the magnetic
field along the $z$ direction to obtain the mean magnetic field $\bar{\boldsymbol{B}}$,
which now depends only on $x$ and $y$. We then use the mean field
to calculate the reconnection rate in the same manner as the calculation
for the 2D reconnection rate. Specifically, because the mean magnetic field $\bar{\boldsymbol{B}}$ is divergence-free, we can construct a mean-field flux function $\bar{\psi}$, subject to the boundary condition $\bar{\psi}=0$ on the conducting-wall boundaries,  such that $\bar{\boldsymbol{B}}=\bar{B_{z}}\boldsymbol{\hat{z}}+\boldsymbol{\hat{z}}\times\nabla\bar{\psi}$. We then identify the dominant X-point of the mean-field flux function and its corresponding value of the flux function $\bar{\psi}_X$. The reconnection rate is determined by $d\bar{\psi}_X/dt$.

\begin{figure}
\includegraphics[width=1\columnwidth]{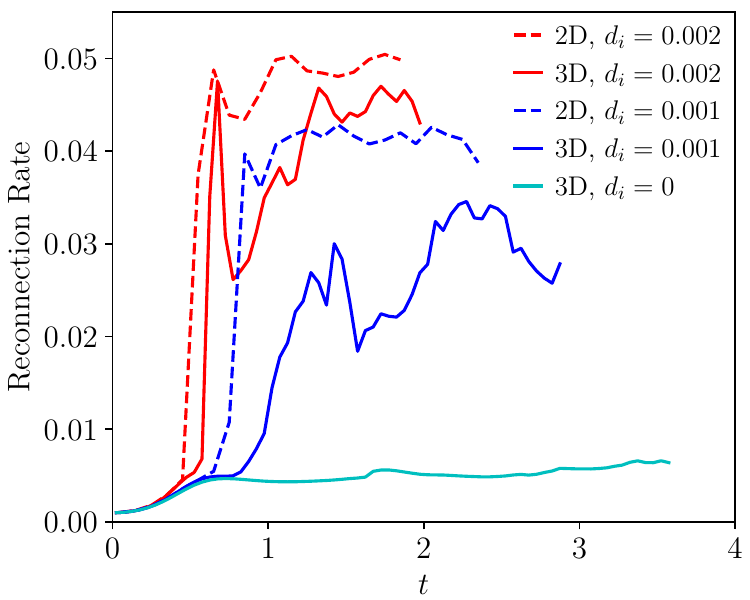}

\caption{Time histories of the reconnection rates for different cases. Solid
lines correspond to 3D simulations, and dashed lines are 2D simulations.\label{fig:Time-history-of-rec-rate}}
\end{figure}

Figure \ref{fig:Time-history-of-rec-rate}
shows the time histories of reconnection rates for these 2D and 3D
runs. The first three cases, all of which end up in a single-X-line
configuration, yield faster reconnection rates. In comparison, reconnection
in the fourth case, which does not settle to a single-X-line configuration, is slower. For reference, we also add a 3D resistive MHD (i.e., $d_{i}=0$)
run, which gives a substantially slower reconnection rate.

\begin{figure}
\includegraphics[width=1\columnwidth]{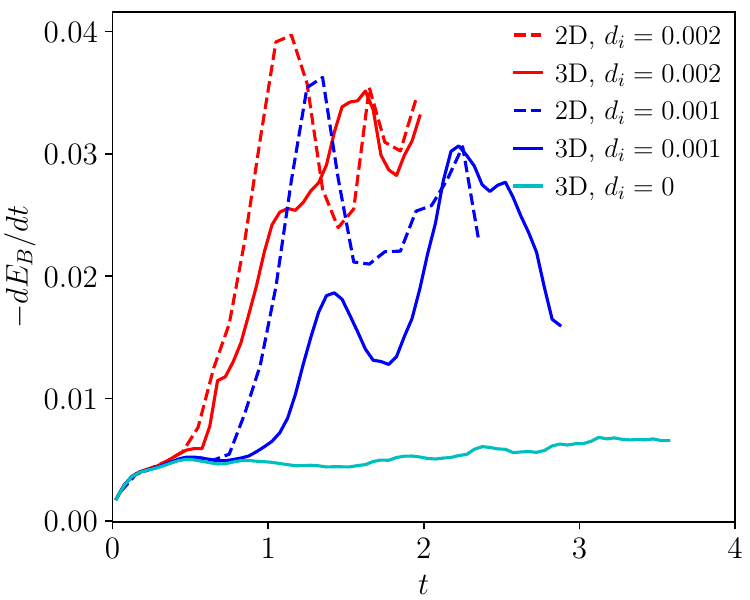}

\caption{Time histories of the total magnetic energy decaying rate for different cases. Solid
lines correspond to 3D simulations, and dashed lines are 2D simulations.\label{fig:Time-history-of-magnetic-energy-decay-rate}}
\end{figure}

Note that properly measuring reconnection rates in complex 3D geometries remains a challenge, for which various methods have been proposed (see, e.g.~ Ref.~[\onlinecite{DaughtonNKRL2014}] and references therein). For background configurations exhibiting translation symmetry, our approach of utilizing the mean magnetic field to calculate the reconnection rate provides a straightforward and effective estimate. The obtained reconnection rates qualitatively align with the trends observed in magnetic energy decay rates, $-dE_B/dt$, which quantify the rate at which magnetic energy is converted into other energy forms. Here, the magnetic energy $E_B=\int B^2/2 d^3x$, integrated over the entire simulation box. Figure \ref{fig:Time-history-of-magnetic-energy-decay-rate} presents the magnetic energy decay rates for the same 2D and 3D simulations as in Figure \ref{fig:Time-history-of-rec-rate}. The results indicate that single-X-line Hall reconnection is more efficient at releasing magnetic energy compared to 3D Hall MHD turbulent reconnection, while resistive MHD exhibits the slowest energy release.

\section{Characteristics of the Self-Generated Turbulent State \label{sec:Characteristics-of-the-Turbulence}}

\begin{figure}
\includegraphics[width=1\columnwidth]{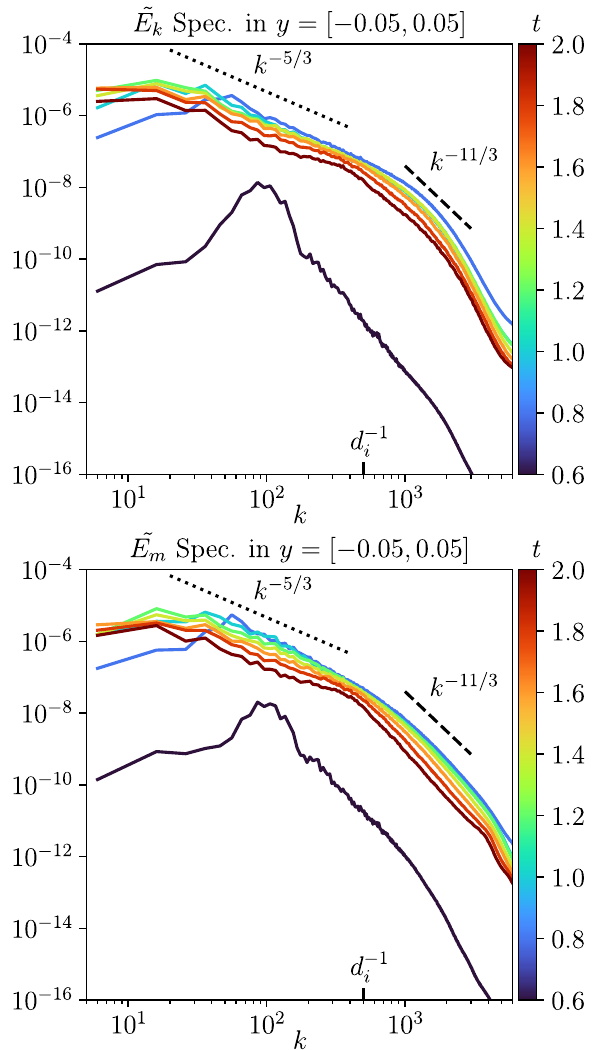}

\caption{Time sequences of the kinetic (upper panel) and the magnetic energy
(lower panel) spectra integrated over the range from $y=-0.05$ to
$0.05$ for the case $d_{i}=0.002$. The tick labels of the colorbar
indicate the times corresponding to the curves.\label{fig:spectra}}
\end{figure}

\begin{figure}
\includegraphics[width=1\columnwidth]{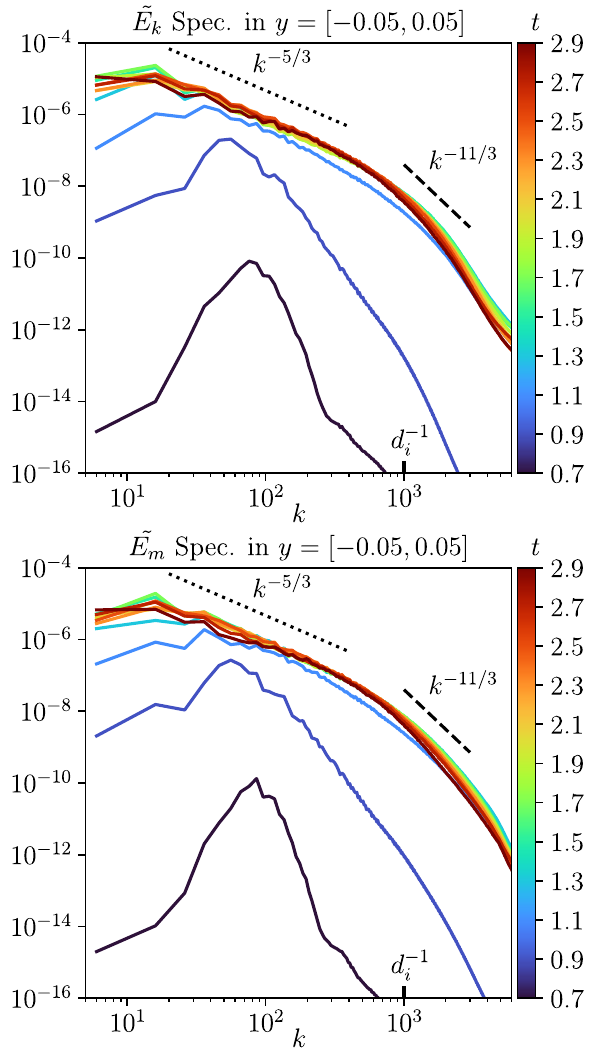}

\caption{Time sequences of the kinetic (upper panel) and the magnetic energy
(lower panel) spectra integrated over the range from $y=-0.05$ to
$0.05$ for the case $d_{i}=0.001$. The tick labels of the colorbar
indicate the times corresponding to the curves.\label{fig:spectra1}}
\end{figure}
\begin{figure*}
\includegraphics[width=1\textwidth]{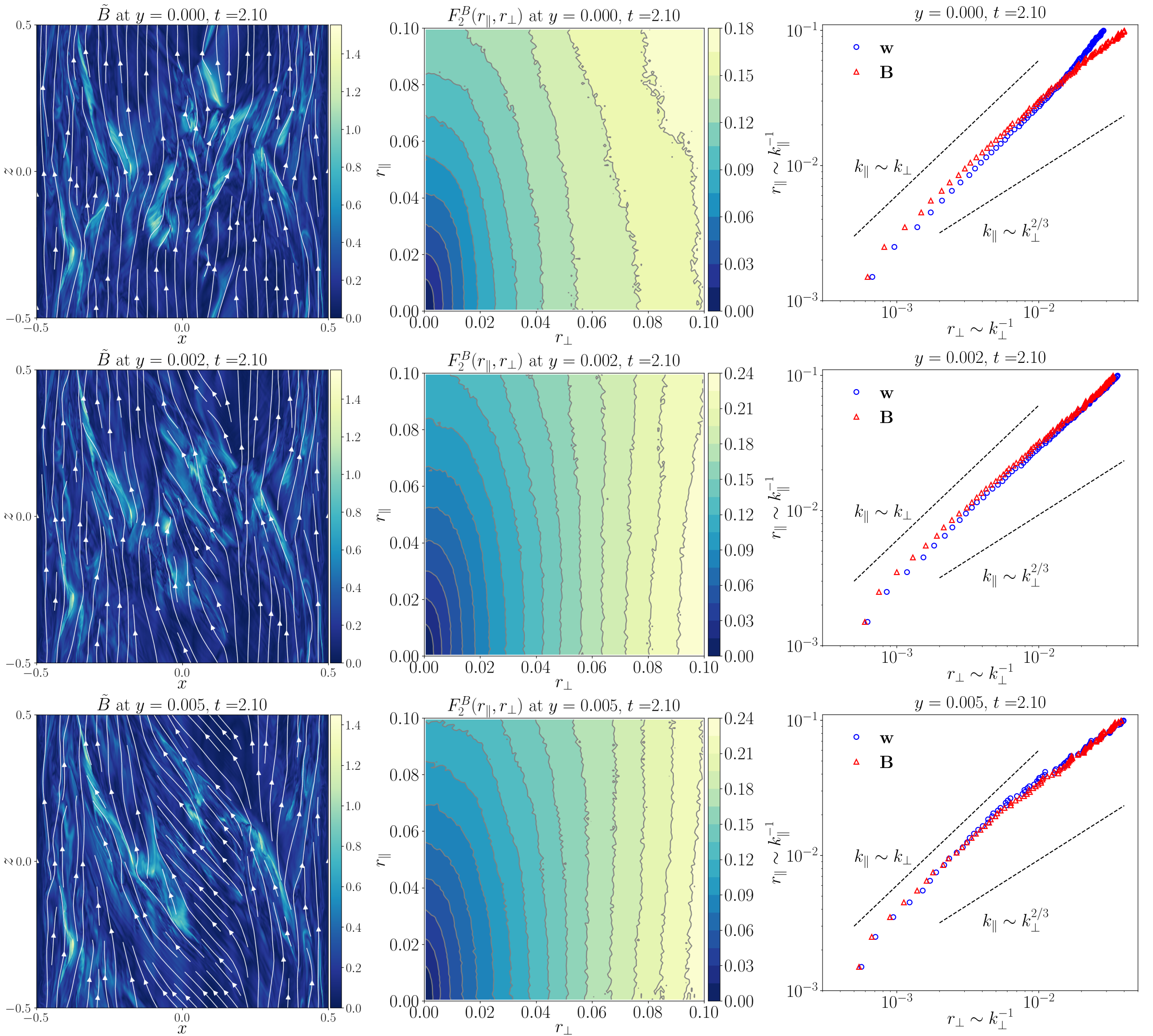}

\caption{Two-point structure function analysis of the fully developed turbulent state for the case with $d_{i}=0.001$ at $t=2.1$. The first column
shows the magnitude of the fluctuating part of the magnetic field
$\tilde{B}$, overlaid with streamlines of the in-plane magnetic field.
The second column shows contours of the structure function $F_{2}^{B}(r_{\parallel},r_{\perp})$.
The third column shows the relations between the semi-major axis $r_{\parallel}\sim k_{\parallel}^{-1}$
and the semi-minor axis $r_{\perp}\sim k_{\perp}^{-1}$ of contours
of the structure functions $F_{2}^{B}(r_{\parallel},r_{\perp})$ and
$F_{2}^{w}(r_{\parallel},r_{\perp})$. These relationships quantify the scale dependence of anisotropy in turbulent eddies. The two dashed lines represent the scalings of $k_{\parallel}\sim k_{\perp}$ (scale-independent)
and $k_{\parallel}\sim k_{\perp}^{2/3}$ (Goldreich--Sridhar theory),
for reference. The first row shows the results at $y=0$, the second row at $y=0.002$, and the third row at $y=0.005$, respectively.
\label{fig:Two-point-structure-functions}}
\end{figure*}

We further investigate whether the reconnection layer evolves to a
self-generated turbulent state in the two 3D Hall MHD simulations
by examining the energy spectra and two-point structure functions
of the kinetic and magnetic fluctuations \emph{within} the layer.
Special treatments are necessary for calculating energy spectra and
structure functions due to the strong inhomogeneity and shear in both magnetic and velocity fields.
We adopt the procedure of Huang and Bhattacharjee (2016), \citep{HuangB2016} which is summarized below.

To calculate the kinetic energy spectrum, we first define a weighted
velocity field $\boldsymbol{w}\equiv\rho^{1/2}\boldsymbol{v}$ such
that the kinetic energy density is a quadratic form $w^{2}/2$. We
then decompose $\boldsymbol{w}$ into the sum of the mean field $\bar{\boldsymbol{w}}$
and the fluctuation $\tilde{\boldsymbol{w}}$, where the mean field is defined as the average along the $z$ direction. The total kinetic energy is the sum of the
mean-field kinetic energy $\bar{E_{k}}\equiv\int\bar{w}^{2}/2\,d^{3}x$
and the fluctuation part of the kinetic energy $\tilde{E_{k}}\equiv\int\tilde{w}^{2}/2\,d^{3}x$.
Likewise, we decompose the magnetic field $\boldsymbol{B}$ into the
mean field $\bar{\boldsymbol{B}}$ and the fluctuation $\tilde{\boldsymbol{B}}$.
The total magnetic energy is the sum of the mean-field magnetic energy
$\bar{E_{m}}\equiv\int\bar{B}^{2}/2\,d^{3}x$ and the fluctuation
part of the magnetic energy $\tilde{E_{m}}\equiv\int\tilde{B}^{2}/2\,d^{3}x$.
We only consider the fluctuation parts for the calculation of the kinetic and magnetic energy spectra. 

Before we calculate the turbulence energy spectra within the reconnection layer, we first multiply the fluctuating fields $\tilde{\boldsymbol{B}}$
and $\tilde{w}$ by a $C^{\infty}$ Planck-taper window function,\citep{McKechanRS2010}
which equals unity within the range $-0.2\le x\le0.2$ and tapers
off smoothly to zero over the ranges where $0.2\le\left|x\right|\le0.4$.
This step reduces the influence of irrelevant fluctuations from downstream exhaust regions, allowing us to focus on turbulence within the reconnection layer. Next, we compute the discrete Fourier energy spectra of $\tilde{E_{k}}$
and $\tilde{E_{m}}$ using the ``windowed'' variables on each constant-$y$
slice to obtain 2D energy spectra as functions of the wave numbers
$k_{x}$ and $k_{z}$. Then, we integrate the 2D energy spectra over
the reconnection inflow direction $y$, which is also the direction
of the strongest inhomogeneity, within the range $-0.05\le y\le0.05$.
Finally, we calculate one-dimensional (1D) spectra as functions of
$k\equiv\sqrt{k_{x}^{2}+k_{z}^{2}}$ by integrating over the azimuthal
direction on the $k_{x}-k_{z}$ plane. 

Figure \ref{fig:spectra} shows the time evolution of the resulting
1D spectra of $\tilde{E_{k}}$ and $\tilde{E_{m}}$ for the case of
$d_{i}=0.002$. Both spectra exhibit qualitatively similar behaviors
and remain close to equipartition between the two. When the plasmoid
instability occurs at about $t=0.6$, the energy spectra peak at $k\simeq100$.
Subsequently, the energy cascades in both forward and inverse directions to smaller and larger scales and quickly fills a broad range of scales at around $t=1.0$. However,
because the reconnection geometry eventually evolves to a single-X-line
configuration, the energy spectra do not settle to a quasi-steady
state. Instead, the fluctuation parts of the energies peak at around
$t=1.0$, then gradually decay as the reconnection geometry becomes
increasingly quasi-2D. 

In comparison, the time evolution of the energy spectra for the case with $d_i=0.001$, shown in Fig.~\ref{fig:spectra1}, is qualitatively similar to the previous case at an early stage. However, because this case does not settle to a single-X-line configuration, the reconnection layer appears to realize a self-generated turbulent state, where the energy spectra become nearly time-independent after $t=1.7$. 

During the quasi-steady phase, the kinetic and magnetic energy spectra
in Figure \ref{fig:spectra} approximately follow the $k^{-5/3}$
power law in the MHD range when $kd_{i}<O(1)$. The spectra steepen
in the sub-$d_{i}$ range when $kd_{i}>O(1)$. As a reference, we
plot the $k^{-11/3}$ power law in the sub-$d_{i}$ range, which is the prediction of Galtier and Buchlin.\citep{GaltierB2007} However,
because we do not have a sufficient separation between the ion skin depth $d_{i}$ and the dissipation scale, the energy spectra do not
exhibit a definitive power law in the sub-$d_{i}$ range. The same two power laws are also shown in Figure \ref{fig:spectra} for reference, even though the energy spectra do not reach a quasi-steady phase. Notably, the energy spectra in this case also steepen in the sub-$d_{i}$ range.

Next, we further investigate the alignment of turbulence eddies with the local magnetic field by calculating two-point structure functions of the kinetic and magnetic fluctuations in terms of the parallel displacement $r_{\parallel}$ and the perpendicular displacement $r_{\perp}$ relative to the local magnetic field;
i.e., $F_{2}^{w}(r_{\parallel},r_{\perp})\equiv\left\langle \left|\boldsymbol{w}\left(\boldsymbol{x}+\boldsymbol{r}\right)-\boldsymbol{w}\left(\boldsymbol{x}\right)\right|^{2}\right\rangle $
and $F_{2}^{B}(r_{\parallel},r_{\perp})\equiv\left\langle \left|\boldsymbol{B}\left(\boldsymbol{x}+\boldsymbol{r}\right)-\boldsymbol{B}\left(\boldsymbol{x}\right)\right|^{2}\right\rangle $.
Structure functions allow us to measure the scale dependence of the anisotropy of turbulence eddies. In MHD turbulence, Goldreich \& Sridhar (GS) theory \citep{GoldreichS1995,GoldreichS1997} predicts that turbulence eddies become increasingly more elongated along the magnetic field at smaller scales. More precisely, GS theory predicts a scale-dependent relationship $k_{\parallel}\sim k_{\perp}^{2/3}$ between the wavenumbers parallel ($k_\parallel$) and perpendicular ($k_\perp$).

The GS theory relies on the critical balance condition, which assumes that the time scale for the nonlinear energy cascade of an eddy in directions perpendicular to the magnetic field balances the Alfv\'en wave propagation time scale of the eddy along the field. However, in Hall MHD, the dispersive whistler waves at sub-$d_i$ scales alter the parallel wave propagation time and may cause different anisotropic scaling relations.\citep{ChoL2004} At MHD scales when $k_\parallel d_i \ll 1$, the whistler waves become the shear Alfv\'en wave, and the GS theory may still be applicable.

The GS scaling relation $k_{\parallel}\sim k_{\perp}^{2/3}$ has been confirmed by Cho and Vishniac for homogeneous MHD turbulence using two-point structure function analysis.\citep{ChoV2000}  In self-generated turbulence within reconnection layers, however, previous resistive MHD studies have shown deviations from the GS scaling relation.\citep{HuangB2016,YangLGLLHZF2020}  Here, we explore whether similar behavior is observed in MHD-scale eddies for Hall MHD turbulent reconnection. In particular, we investigate whether our simulations of inhomogeneous reconnection layers in Hall MHD also exhibit deviations from the GS scaling relation.

Here we adopt the procedure of Huang and Bhattacharjee (2016),\citep{HuangB2016} which is a modification of the procedure of Cho and Vishniac (2000)\citep{ChoV2000} to accommodate the inhomogeneous background of the reconnection layer. Due to the strong inhomogeneity along the $y$ direction in our system, we calculate the structure functions for each $x$--$z$ plane, restricting displacements to lie within these individual planes, rather than using the full 3D space. For a pair of points, we define the local magnetic field as the average of the magnetic fields from the two points. To ensure consistency with displacements limited to individual $x$--$z$ planes, we measure the parallel and perpendicular components of the displacement relative to the in-plane component of the local magnetic field. We compute the structure functions by averaging over $10^{9}$ random pairs of points; the $x$ coordinates
of these points are limited to the range $-0.25\le x\le0.25$.

We focus on the case with $d_{i}=0.001$ because it appears to realize
a self-generated turbulent state in the reconnection layer. In the
following discussion, we present the results from the snapshot at
$t=2.1$, but other snapshots during the quasi-steady period show
similar behaviors. Figure \ref{fig:Two-point-structure-functions}
summarizes the outcome of this analysis. Here, the rows correspond
to different $x$-$z$ slices. The first row shows the results at
$y=0$, the second row at $y=0.002$, and the third row at $y=0.005$,
respectively. The first column of the figure shows the magnitude of
the magnetic field fluctuation $\tilde{B}$, overlaid with streamlines
of the in-plane magnetic field. The magnetic field fluctuations form
elongated eddies along the direction of the local magnetic field.
The second column shows contours of the structure function $F_{2}^{B}(r_{\parallel},r_{\perp})$,
which clearly show elongated eddies along the local magnetic
field direction. We have also performed the same analysis for the
weighted velocity field fluctuation $\tilde{w}$ and the corresponding
structure function $F_{2}^{w}(r_{\parallel},r_{\perp})$. The results
are qualitatively similar to those of the magnetic field fluctuations; therefore, they are not shown. The third column shows the relations between the semi-major axes $r_{\parallel}\sim k_{\parallel}^{-1}$ and the semi-minor axes $r_{\perp}\sim k_{\perp}^{-1}$ of contours of the
structure functions $F_{2}^{B}(r_{\parallel},r_{\perp})$ and $F_{2}^{w}(r_{\parallel},r_{\perp})$.
These relations quantify the scale dependence of turbulent eddy anisotropy.
The two dashed lines represent the scalings of $k_{\parallel}\sim k_{\perp}$
(scale-independent) and $k_{\parallel}\sim k_{\perp}^{2/3}$ (GS theory),
for reference. In the first two rows at $y=0$ and
$y=0.002$, relatively close to the mid-plane, the relations between
$r_{\parallel}$ and $r_{\perp}$ appear to be approximately scale-independent. This result is similar to previous resistive MHD results.\citep{HuangB2016,YangLGLLHZF2020}  Interestingly, as we move further away from the mid-plane to $y=0.005$ (last row), the GS scaling relation $k_{\parallel}\sim k_{\perp}^{2/3}$ is partially recovered and appears to be consistent with eddies much larger than the $d_i$ scale.

\section{Discussion and Conclusion\label{sec:Conclusion}}

In conclusion, our simulations show that the occurrence of single-X-line Hall reconnection in 3D is less probable compared to 2D. A case that evolves to a single-X-line geometry can become turbulent with multiple reconnection sites in 3D. Notably, the single-X-line geometry consistently yielded the highest reconnection rate among all studied cases.

Whether  3D Hall MHD  reconnection realizes a self-generated turbulent state or a single-X-line configuration depends on the interplay of plasma parameters, including the Lundquist number $S$ and the system size to kinetic scale ratio $\Lambda=L/d_{i}$. This study shows that at a fixed Lundquist number $S=2\times10^5$, a system of $\Lambda=500$ evolves to a single-X-line geometry, whereas a system of a larger system size $\Lambda=1000$ evolves to a turbulent reconnection state. However, whether the system with $\Lambda=1000$ can lead to a single-X-line configuration in even higher Lundquist numbers remains an open question. To comprehensively understand the diverse dynamic behaviors of reconnection across parameter space, a systematic investigation is imperative. Such research will be crucial in advancing our knowledge of reconnection processes in complex plasma systems.

The self-generated turbulent reconnection in 3D Hall MHD exhibits significant qualitative differences compared to resistive MHD. The turbulent reconnection layer in Hall MHD is significantly broader, and the reconnection rate is much faster. Regardless of whether the reconnection geometry settles to a single X-line or becomes turbulent, the hallmarks of Hall reconnection remain visible. Notably, the reconnection exhaust regions open up and form a Petschek-like configuration (see Figures \ref{fig:3D_time_seq_1} and \ref{fig:3D_time_seq_2}). We emphasize that the resemblance with the Petschek-like configuration is just geometrical and nothing more because the underlying Hall MHD dynamics is qualitatively different from the resistive MHD dynamics. Nonetheless, our simulation results and future research for even larger system sizes may shed light on recent Parker Solar Probe in situ measurements of reconnection exhausts associated with interplanetary coronal mass ejections and heliospheric current sheets, where bifurcated current sheets resembling Petschek's reconnection model with a pair of slow-shocks or rotational discontinuities have been reported. \citep{PhanBE2020}

Many questions remain open, particularly for applications to natural reconnection events where the system sizes are much larger than those considered in this study (i.e., $\Lambda=L/d_{i}\ggg1$). Will a larger system size make self-generated turbulent reconnection easier to realize? Do global reconnection rate and geometry depend on the microscopic description of reconnection physics (e.g., Hall MHD or fully kinetic PIC) when the scale separation between $L$ and $d_{i}$ is large? In the future, the Hall MHD description of self-generated turbulent reconnection should be further compared with fully-kinetic PIC simulations or more elaborated fluid models such as high-moment multi-fluid models,\citep{WangHBG2015,NgHHBSDWG2015} in particular in the regime when the system size is much larger than kinetic scales. 

The self-generated turbulence in the reconnection layer also poses new challenges to turbulence theories and future numerical simulations. Our Hall simulation shows that the turbulent state in the vicinity of the reconnection mid-plane does not satisfy the Goldreich-Sridhar (GS) scaling relation of turbulence eddy anisotropy. This result is similar to previous findings in resistive MHD simulations. However, away from the mid-plane, the GS scaling relation is partially recovered for large eddies. This phenomenon has not been observed in resistive MHD simulations, and is not predicted by any theory. 

Why the GS scaling relation is not satisfied in resistive MHD self-generated turbulent reconnection but is partially recovered in Hall MHD remains an open question.  Previously, Huang and Bhattacharjee argued that the discrepancy may be because the reconnection layer is highly inhomogeneous, with the reconnection outflow and the magnetic field being strongly sheared.\citep{HuangB2016} In Hall MHD turbulent reconnection, the turbulent region is significantly thicker than the resistive MHD counterpart. Because the magnetic field is not as strongly sheared at planes away from the mid-plane, this observation could explain why the GS scaling relation is partially recovered in places away from the mid-plane. 

Even within the MHD regime, whether self-generated turbulent reconnection deviates from the GS theory has been unsettled. In a previous study, Kowal \textit{et al.}~showed that the scaling relation for large-scale eddies in turbulent reconnection simulations satisfies the GS scaling, while small-scale eddies are ``contaminated'' by reconnection and exhibit different scaling behaviors.\citep{KowalFLV2017} However, the initial and boundary conditions in their study differ from those in the present study: their initial condition contains a tangential discontinuity rather than a smooth magnetic field; the strength of the guide field is considerably lower than that of the reconnecting component of the magnetic field; the boundary conditions are periodic in the outflow direction and open in the inflow direction. It is unclear how much of their findings can be ascribed to these distinctions.

Furthermore, Kowal \textit{et al.} employed full 3D fields and displacements in their calculation of two-point structure functions, while our analysis is restricted to displacements within each $x$-$z$ slice, utilizing in-plane magnetic field components for determining parallel and perpendicular displacements. The implications of this significant difference warrants further investigation. Our decision to use 2D slices stems from the pronounced inhomogeneity along the inflow ($y$) direction, which renders displacements in this direction qualitatively distinct from in-plane displacements. Preliminary attempts to calculate two-point structure functions using full 3D fields and displacements yielded results that were challenging to interpret. Future simulations with higher resolution, allowing for sufficient scale separation between the turbulent region thickness and the smallest turbulence eddies, should revisit this analysis to comprehensively assess the consequences of employing 2D versus 3D displacements.

It is worth mentioning that the present study uses a compressible code. The relatively high plasma $\beta$ ($\approx 4$ relative to the reconnecting component of the magnetic field) and the presence of a guide field keep the plasma close to, but not perfectly, incompressible. Throughout the simulations, the root-mean-square density fluctuation is approximately $3\%$, although the density can differ from the mean value by up to $10\%$ at some locations. Even though the compressibility of the plasma is relatively low, the compressible waves within the system may introduce subtle effects on the energy cascade and eddy anisotropy.\citep{Galtier2023}  

In order to assess the effects of compressibility, turbulent reconnection should be investigated with incompressible MHD and Hall MHD, as well as compressible systems with varying plasma $\beta$. Furthermore, simulations with varying guide field strengths will clarify magnetic shear effects. In a broader context, it is also of great interest to further investigate how Hall and kinetic effects affect plasmoid-mediated energy cascade in homogeneous turbulence, which is an important topic of current research.\citep{BoldyrevL2017,MalletSC2017,ComissoHLHB2018,DongWHCB2018,DongWHCSB2022,CerriPLSK2022} 

\section{Supplementary Material}

See the supplementary material for an animation for the 2D simulation with $d_{i}=0.001$ in the reconnection layer, similar to that shown in Figure \ref{fig:2D_time_seq}.

\begin{acknowledgments}
This research was supported by the U.S. Department of Energy, grant
number DE-SC0021205, the National Science Foundation, grant numbers AGS-2301337 and 2209471, and the National Aeronautics and Space Administration, grant number 80NSSC18K1285. Computations were performed on facilities at the National Energy Research Scientific Computing Center and the National Center for Atmospheric Research.
\end{acknowledgments}

\bibliographystyle{apsrev4-2}
\bibliography{Hall_Turb_Rec}

\end{document}